# A New Framework of Authentication Over Cloud Computing

Megouache Leila[1], Zitouni Abdelhafid[1], and Djoudi Mahieddine[2]

[1] Lire Labs, Abdelhamid Mehri Constantine 2 University, Ali Mendjli,
25000 Constanine, Algeria
`{leila.megouache,
abdelhafid.zitouni}@univ-constantine2.dz`
[2] TECHNE Labs, University of Poitiers, 1 rue Raymond Cantel,
86073 Poitiers Cedex 9, France
`mahieddine.djoudi@univ-poitiers.fr`

**Abstract.** The growth of local data annually implies extra charges for the customers, which makes their business slowing down. Cloud computing paradigm comes with new technologies that offer a very economic and cost-effective solution, but the organizations are slow in accepting it due to security issues and challenges associated with it. Security is one of the major issues which hamper the growth of cloud and it is highly crucial to protect the sensitive data and systems in the cloud setting in order to ensure the privacy of its users. This work explores the new directions in cloud computing security, while highlighting the correct selection of these fundamental technologies from cryptographic point of view.

## 1 Introduction

Cloud computing is becoming increasingly a magical solution and a widely adopted technology for delivering services over the Internet, such as infrastructure, platform or software with a reasonable and more and more decreasing cost for the users. However, cloud computing is still in its initial stage [1]. The lack of standards, the security and the interoperability issues hamper the growth of cloud computing [2, 3]. Thus, the choice of clouds made by companies is usually based on the quality of services, but measuring the quality of cloud providers' approach to security is difficult because many cloud providers will not expose their infrastructure to customers.

Therefore, the security is an important factor that should be taken into account by cloud service providers.

Before making the transfer of the data towards Cloud, the company owes classify them and choose Cloud adapted according to the following categories.

For that, the organizations and individuals are very anxious about how security can be guaranteed in the new cloud environment. Moreover, enterprises and individuals have strong constraints on hosting their sensitive data and critical applications on clouds.

So, the biggest challenge in cloud computing is to successfully address the security issues associated with their deployment [4, 5], and to provide evidence to their customers that their data are safe.

To date, there is minimal work done in the field of security of data as compared to traditional data storage. In order to alleviate these security fears, and to add a brick to cloud computing security research, we propose in this paper a new architecture to implement security. It is based on the creation of a virtual private network for ensure the security in the network and encrypt authentication before transferring in the cloud.

This paper is organized as follows: Sect. 2 introduces cloud computing overview. In Sect. 3, we explain security and identification of threats, In Sect. 4, the related works is presented. In Sect. 5, we propose our framework. Finally, In Sect. 6, the conclusions of this work are presented.

## 2 Cloud Computing Overview

Then we present firstly an overview of cloud computing.

A. *Types of cloud computing*

   Considering the installation of network infrastructure a cloud environment can be broadly categorized into three types-public cloud, private cloud and hybrid cloud [7].
   (a) **Public cloud:** the service is provided by a third party via the internet. The physical infrastructure is owned and managed by the service provider. This cloud is less secure compared to other models [4], since all applications and data are available to the public and accessible via the Internet.
   (b) **Private cloud:** it is a dedicated cloud that is managed internally or by a third-party and can be hosted generally on premises or even externally [5]. The physical infrastructure is exclusively used by one organization. This Cloud offer a higher degree of security since only users in the organization have access to the private cloud.
   (c) **Community cloud**: the physical infrastructure is controlled and shared by several organizations and is based on a community of interest [6].
   (d) **Hybrid cloud:** combine two or more distinct cloud infrastructures. This cloud must be linked by a standard technology for data and applications portability.

B. *Cloud Computing Service Architecture*

   Mainly, three types of services you can get from a cloud service provider [8].
   1. Infrastructure as a service- service provider bears all the cost of servers, networking equipment, storage, and back-ups. You just have to pay to take the computing service. And the users build their own application softwares. Amazon EC2 is a great example of this type of service.
   2. Platform as a service-service provider only provide platform or a stack of solutions for your users. It helps users saving investment on hardware and software.
   3. Software as a service- service provider will give your users the service of using their software, especially any type of applications software (Fig. 1).

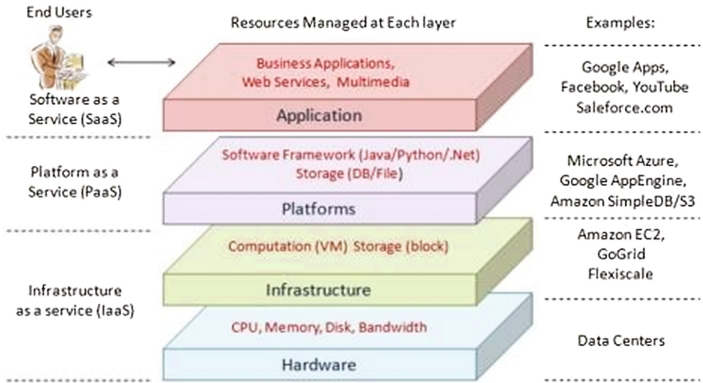

**Fig. 1.** Cloud computing service architect

C. *The main characteristics of Cloud computing are* [6]:
  (a) *On-demand self-service*: users can access and control their services automatically without the intervention of the service provider [6].
  (b) *Broad network access:* Services are available over the internet, they accessible from anywhere regardless of the device used.
  (c) *Resource pooling:* the computing resources are pooled to serve multiple consumers, with different physical and virtual resources [6].
  (d) *Rapid elasticity:* The computing resources are rapidly and dynamically provisioned. Users can increase and decrease the functionality available according to their needs. The capacities appear to be unlimited.
  (e) *Measured service:* The resource use is controlled by leveraged metering capability.

## 3 Security Identification of Treats

Cloud computing due to its architectural design and characteristics imposes a number of security benefits, which include centralization of security, data and process segmentation, redundancy and high availability [7].

A. *Information Security*
  Security in general, is related to the important aspects of confidentiality, integrity and availability; they thus become building blocks to be used in designing secure systems. The basic security issues are: authentication, authorization, auditing, confidentiality and privacy, integrity, availability and non-repudiation [8, 9]:

*Authentication*
Authentication is the process by which a user or other entity's identity is verified.

*Authorization*

Authorization is the process that ensures that an entity has the right credentials to access certain resources or to perform a requested task. This decision is made after authenticating the identity in question. And each entity has a specific level of authorization.

**Confidentiality and privacy**

Confidentiality refers to only authorized parties or systems having the ability to access protected data. The threat of data compromise increases in the cloud, due to the increased number of parties, devices and applications involved, that leads to an increase in the number of points of access.

**Integrity**

Data Integrity refers to protecting data from unauthorized deletion, modification or fabrication. Managing an entity's admittance and rights to specific enterprise resources ensures that valuable data and services are not abused, misappropriated or stolen. By preventing unauthorized access, organizations can achieve greater confidence in data and system integrity.

**Non-repudiation**

Non-repudiation is the ability to limit entities from refuting that a legitimate transaction took place, and to ensure that it has not been changed by a malicious entity.

**Availability**

Availability refers to the property of a system being accessible and usable upon demand by an authorized entity. System availability includes a systems ability to carry on operations even when some authorities misbehave. The system must have the ability to continue operations even in the possibility of a security breach.

**Auditing**

Is a process of recording of the performed tasks and collecting information about user attempting access to a particular resource [8].

B. *Cloud Computing Threats*

Before deciding to migrate to the cloud, we have to look at the cloud security vulnerabilities and threats to determine whether the cloud service is worth the risk due to the many advantages it provides [10].

(a) **Ease of Use:** The cloud services can easily be used by malicious attackers, since a registration process is very simple, because we only have to have a valid credit card.
(b) **Secure Data Transmission:** When transferring the data from clients to the cloud, the data needs to be transferred by using an encrypted secure communication.
(c) **Insecure APIs:** Various cloud services on the Internet are exposed by application programming interfaces. Since the APIs are accessible from anywhere on the Internet, malicious attackers can use them to compromise the confidentiality and integrity of the enterprise customers.
(d) **Malicious Insiders:** Employees working at cloud service provider could have complete access to the company resources.

(e) **Shared Technology Issues:** The cloud service SaaS/PasS/IaaS providers use scalable infrastructure to support multiple tenants which share the underlying infrastructure.

## 4 Related Works

There are a number of work concerning the security, the privacy and the authentication of companies data in the cloud computing.

In [8], the privacy information is encrypted using symmetric encryption algorithm with encryption keys, which are generated from an encryption tree. And then the secret keys are encrypted using public key cryptosystem to send them to receivers. Receivers are able to decrypt the privacy information correctly. However, the cloud storage provider and any illegal users cannot read the privacy information. The analysis shows that the proposed system protects privacy efficiently with no extra storage consumptions.

In [11], proposes a new framework which ensures the data security and integrity and focuses on the encryption and decryption approach facilitating the cloud user with data security assurance. The proposed solution talks about the increased security only but does not talk about the performance. The solution also includes the functioning of forensic virtual machine, malware detection and real time monitoring of the system.

In [12], proposed key-policy attribute based encryption (KP-ABE) scheme where the data owner build the access structure in the user's key and cipher the data using a set of attributes. This scheme was able to achieve the fine grained access control while having more flexibility to control the user's privileges. One of the main problems of this scheme, is that the revocation process includes an update to all the keys of the users.

In [13], Provides an overview of Common approaches to preserve confidentiality in e-Health Cloud. These approaches are classified into two categories:

Cryptographic approaches (based on encryption techniques) and non-cryptographic approaches (mainly use on access control). They also point out the advantages and disadvantages of each approach.

In [14], this, the authors present a contribution to protecting the privacy of Web users. The objective of this work is to allow a client to query the search engine in a way to preserve privacy. This means that the search engine, which receives the request, or any opponent who listens to the network, cannot deduce the identity of the applicant (the user). The authors aim to generate false application (Fake query) that cannot be identified by the opponents (or engine research).

A major disadvantage is the introduction of irrelevant answers to protect the applications in this solution because they added noise to the search request to perform obfuscation (interference). This solution decreases the precision of the results and causes overload on the network.

# 5 Proposed Security Model

Major concerns and issues in security have been discussed in the previous sections. It has been observed that, despite quality research on security data outsourcing and data services for almost a decade, existing approaches on database encryption, certification, digital signatures [12], contractual agreements etc. have not gained much success in operations.

Our proposal framework be to create firstly, one network virtual deprived (VPN) between the customer and the provider of such goes out that the customer little to reach his space Cloud in a secure way. Secondly, using a symmetric cryptographic of data, can offer the efficiency of symmetric cryptography while maintaining the security.

A. *Create a Virtual private network*

In this step we create a virtual private network for ensure the security in the network and the data can be transferred in security.

When users try to connect to VPN client, it will prompt for user and password. After this information is entered and the user, click OK, the client will try to connect with a message, "contacting the security gateway" in the bottom status bar and the connection will timeout. When this happens it will prevent everyone from connecting with VPN client.

The connecting take around 30 s to attempt to connect to the security gateway. Sometimes they ask for a username and password and then fail, or on other times they will just fail directly.

The error message is always "Secure VPN Connection terminated locally by the client, On the other, there is no issue and the connect and prompt for authentication almost instantly (Fig. 2).

The algorithm used to connect vpn customer:

```
Connect "vpncustomer"
begin
Time=0;
Input ( user, password)
        User= information
        Password= *********
"contacting the security gate way"
If time> 30  {
"the connection is fail"
Else
Connecting succefuly; }
End;
```

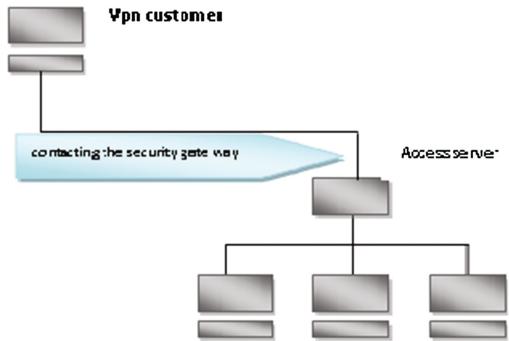

**Fig. 2.** Contacting the security gate

B. *Access with authentication*

When the connection will be establishes, the customer goes on the internet page and opens him on the address URL, Which is delivered by provider, when the page opens another user and password are required. The password is only encrypted by the first client, if another program or malicious program wants to access it, the system will find the passwords encrypt, what it will prevent it from accessing (Fig. 3).

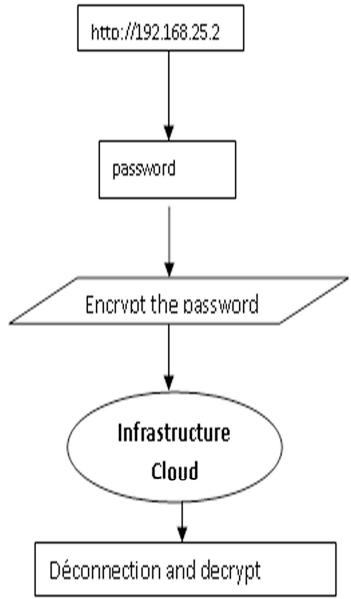

**Fig. 3.** Prototype of authentication

- *Symmetric cryptographic of data*
  In this study we will use the algorithm of Symmetric Encryption Algorithm. Advanced Encryption Standard or *AES* (**S**ymmetric **E**ncryption **A**lgorithm) is a symmetric encryption algorithm. He won in October 2000 the AES competition, launched in 1997 by the NIST and became the new encryption standard for US Government organizations.

The algorithm used to encrypt data:

```
Input   :  table  T  and  key  K
Output  :  table  T  modified
Function AES (T, K)
  Begin
  KeyExpansion (K, TK);
  AddRoundKey (T, TK [0];
  for (i = 1; i<nr; i + +)
  Round (T, TK [i]);
  FInalRound (T, TK [nr]);
  end
```

The algorithm used to decrypt data:
Decryption with AES:
The encryption routine can be reversed and rearranged to produce a decryption algorithm.

```
AES_Decrypt(T, K) {
KeyExpansion(K, RoundKeys); /*
Initial addition */
AddRoundKey(State, RoundKeys[Nr]);
for (r=Nr-1; i>0; r--) {
InvShiftRows(T);
InvSubBytes(T); AddRoundKey(T,
RoundKeys[r]); InvMixColumns(T);

}
/* FinalRound */ InvShiftRows(Out);
 InvSubBytes(Out);
 AddRoundKey(Out,RoundKeys[0]);
}
```

By this solution only one user can be access to services and, it's better to encrypted all the data which will be transfer in the cloud, it sets a lot of time.

## 6 Conclusion and Future Works

There are many traditional countermeasures to mitigate security and authentication issues. Indeed cloud environment is widely distributed, highly dynamic and more threaten by attacks. Therefore, these countermeasures must be improved or changed to work effectively in this type of environment. This paper discussed cloud computing characteristics and focused on the security issues, which can be solved by our prototype of authentication. Further research could be realized to improve and to extend the present work by including and resolving the interoperability issues in the cloud.

## References

1. Baruah, K.C.H., Banerjee, S., Dutta, M.P., Bhunia, C.T.: An improved biometric-based multi server authentication scheme using smart card. Int. J. Secur. Appl. **9**(1), 397–408 (2015)
2. Clement, T.: The security in the cloud computing. Presented in CQSI, October 2012
3. Nithiavathy, R.: Data Integrity and Data Dynamics with Secure Storage Service in Cloud, Proceedings (2013)
4. Banerjee, S., Dutta, M.P., Bhunia, C.T.: An improved smart card based anonymous multi-server remote user authentication scheme. Int. J. Smart Home **9**(5), 11–22 (2015)
5. Kumbhare, A., Simmhan, P., Prasanna, V.: Cryptonite: a secure and performant data. In: IEEE 5th International Conference on Cloud Computing (CLOUD), pp. 510–519. IEEE (2016)
6. Huang, H., Liu, K.: Efficient key management for preserving HIPAA regulations. J. Syst. Softw. **84**, 122–221 (2015)
7. El Makkaoui, K., Ezzati, A., Beni-Hssane, A., Motamed, C.: Data confidentiality in the word of cloud. J. Theor. Appl. Inf. Technol. **84**(3) (2016)
8. Pippal, R.S., Wu, S.: Robust smart card authentication scheme for multi-server architecture. Wirel. Pers. Commun. **72**(1), 729–745 (2013)
9. Idrissi, H.K., Kartit, A., El Marraki, M.: A taxonomy and survey of cloud computing. Presented at the Security Days (JNS3), pp. 1–5 (2013)
10. Merkle, R.C.: A certified digital signature. In: Proceedings of 9th Annual International Cryptology Conference on Advances in Cryptology, CRYPTO 1989, Santa Barbara, California, USA, vol. 435, pp. 218–238 (1989)
11. Sirohi, P., Agarwal, A.: Cloud computing data storage security framework relating to data integrity, privacy and trust, pp. 4–5 (2015)
12. Sun, X., Qu, S., Zhu, X., Zhang, M., Ren, Z., Yang, C.: Cloud storage architecture achieving privacy protection and sharing. Appl. Math. Inf. Sci. **9**(3), 1639–1644 (2015)
13. Deyan, D.C., Zhao, H.: Data security and privacy protection issues in CC. In: IEEE International Conference on Computer Science and Electronics Engineering (ICCSEE) (2015)
14. Petit, A., Ben Mokhtar, L.B., Kosch, H.: Towards efficient and accurate privacy preserving web search. In: MW4NG 2014, Bordeaux, France, 8–12 December (2014)